 \documentclass{emulateapj}

\usepackage{graphicx}
\usepackage{natbib}
\usepackage{amsmath}
\usepackage{wasysym}
\usepackage{hyperref}
\usepackage{breakurl}
\usepackage{footmisc}
\usepackage{longtable}
\usepackage{multirow}
\usepackage[normalem]{ulem}
\usepackage{bm}

\shorttitle{Photometric study of fourteen low-mass binaries}
\shortauthors{Korda et al.}

\begin{document}
\title{Photometric study of fourteen low-mass binaries\thanks{Based on observations collected via 65-cm telescope at the Ond\v{r}ejov observatory in Czech Republic.}}

   \author{D. Korda\altaffilmark{1}
          \and
          P. Zasche\altaffilmark{1}
          \and
          M. Wolf\altaffilmark{1}
          \and
          H. Ku\v{c}\'{a}kov\'{a}\altaffilmark{1}
          \and
          K. Ho\v{n}kov\'a\altaffilmark{2}
          \and
          J. Vra\v{s}til\altaffilmark{1}
          }

 \affil{
  \altaffilmark{1} Astronomical Institute, Charles University, Faculty of Mathematics and Physics, CZ-180~00, Praha 8, \\
             V~Hole\v{s}ovi\v{c}k\'ach 2, Czech Republic\\
              \email{d.korda@sirrah.troja.mff.cuni.cz}
  \altaffilmark{2}
             Variable Star and Exoplanet Section of Czech Astronomical Society, Vset\'{\i}nsk\'a 941/78, CZ-757 01, Vala\v{s}sk\'e Mezi\v{r}\'{\i}\v{c}\'{\i}, Czech Republic  \\
 }

\begin{abstract}
\noindent
New CCD photometric observations of fourteen short-period low-mass eclipsing binaries (LMB) in the photometric filters I, R and V were used for the light curve analysis. There still exists a discrepancy between radii as observed and those derived from the theoretical modelling for LMB in general. Mass calibration of all observed LMB was done using only the photometric indices. The light curve modelling of these selected systems were performed, yielding the new derived masses and radii for both components. We compared these systems with the compilation of other known double-lined LMB systems with uncertainties of masses and radii less then 5 \%, which includes 66 components of binaries where both spectroscopy and photometry were combined together. All of our systems are circular short-period binaries, and for some of them the photospheric spots were also used. A purely photometric study of the light curves without spectroscopy seems unable to achieve high enough precision and accuracy in the masses and radii to provide for a meaningful test of the M-R relation for low-mass stars.
\end{abstract}


%-------------------------------------------------------------------

\section{Introduction}
Low-mass stars \citep[$M \lesssim 1.0\,M_{\astrosun}$,][]{Morales_2011} are the most common type of stars \citep{Kroupa_2013}. Despite this fact, it seems that models describing the evolution of the stars are not sufficiently robust and show inconsistencies around 3 \% in radius at a fixed mass and up to 10 \% in some cases \citep{Feiden_2012, Spada_2013}. This can be explained by the presence of large-scale magnetic fields that reduce convective efficiency and induce stellar spots on the surfaces of the stars (see e.g. \citet{Chabrier_2007} or \citeauthor{Torres_2010} (\citeyear{Torres_2010}). These phenomena are not often included in models of stellar evolution as well as the rotation of stars. Inconsistency appears to be more apparent for close binary systems due to the high rotation velocity. Rotation period is synchronized with the orbital period due to the tidal forces.

To date, over sixty of such components were observed both via spectroscopy and photometry together and analysed simultaneously, yielding the precise and accurate (uncertainties of masses and radii less than 5 \%) values of masses and radii. The main motivation of this work was the easy availability of a large number of high-precision photometric observations (for example from the Kepler survey), in which a large number of LMB was detected \citep{Kepler}. Therefore, we decided to test whether the M-R relation discrepancy could be observed also from the sole ground-based photometry.

\section{Data}
All new photometric observations of fourteen binaries were carried out in the Ond\v{r}ejov Observatory in the Czech Republic with the 0.65-m~reflecting-type telescope and the G2-3200 CCD camera. Observations were collected from Feb 2015 to Nov~2016 in filters I, R and V \citep{Bessell}. Some of the older observations obtained only in R filter were used for refining the individual orbital periods.

The stars were chosen mainly from the catalog of \citet{Hoffman_2008}. For the selection of suitable stars we used several criteria. Its classification as a low-mass binary was done using the photometric indices $J-H$ and $H-K$ which are known from the 2MASS survey \citep{2MASS_2003} ($J-H > 0.25$ and $H-K > 0.07$ \citet{Pecaut_2013}\footnote{\url{www.pas.rochester.edu/~emamajek/EEM_dwarf_UBVIJHK_colors_Teff.txt}}). Furthermore, we selected binary systems which have short orbital periods ($P<1.5$ days). There were two reasons for this: First, we wanted to analyse as many systems as possible, and secondly, discrepancy should be more apparent for the short-period systems. We did not consider observing binaries with longer orbital periods. Furthermore, we chose the declination to be higher than $+$30$^{\circ}$. This guaranteed that the systems are easily observable throughout most of the year from our observatory. The last criterion was that these systems were not analysed in detail before. We chose eleven systems in Hoffman's catalog, two more were found in measured field (one of them is on the edge of criteria) and one star was added later.

An example of the individual data are listed in Table \ref{tab:data}. Some more details about the observations are listed in Table \ref{tab:summary}.

\begin{table}
	\caption{A sample of photometric observations for NSVS 363024, filter I.}
	\label{tab:data}
	\centering
    \begin{tabular}{c c c}
	\hline \hline
	$H\!J\!D$ & $\Delta \mathrm{m}$ & $\sigma\left(\Delta \mathrm{m}\right)$\\
	$[-2400000]$ & $[\mathrm{mag}]$ & $[\mathrm{mag}]$\\
	\hline
	57262.2917 & 0.6499 & 0.0076\\
	57262.2930 & 0.6657 & 0.0070\\
	57262.2946 & 0.6580 & 0.0065\\
	57262.2955 & 0.6541 & 0.0052\\
	57262.2968 & 0.6605 & 0.0090\\
	57262.2981 & 0.634 & 0.013\\
	57262.2994 & 0.6296 & 0.0075\\
	57262.3033 & 0.6451 & 0.0044\\
	57262.3052 & 0.6577 & 0.0043\\
	57262.3065 & 0.6582 & 0.0044\\
	\hline
	\multicolumn{3}{l}{All photometric observations are available on-line.}
	\end{tabular}
\end{table}

\section{Analysis}
Standard calibration (dark frame, flat field) was applied to the observed frames. The light curves were obtained using an aperture photometry \citep{Motl}\footnote{\url{http://c-munipack.sourceforge.net/}}. As a comparison star we chose a star which satisfies (as best as possible) the following conditions. That star should be as bright as the binary. And spectral type of comparison and binary must be the same. Of course, comparison's brightness is constant.

The light curves were analysed using the program PHOEBE \citep{Phoebe}, which is based on the Wilson-Devinney algorithm \citep{W-D}. The temperature of the primary component was (similarly to mass) calibrated from the above mentioned tables. Orbital periods were known with only a small accuracy from the Hoffman's publication. Better precision was achieved using the methods of minimizing the phase dispersion and construction of the $O-C$ diagrams (used minima are in Table \ref{tab:minima}). PHOEBE is not designed for determining the orbital period.

No spectroscopic observations were obtained for our selected systems. To estimate the individual component masses we assumed relation between the infrared photometric indices $J-H$ and $H-K$ and mass\footnote{The reddening due to interstellar extinction is small for infrared indices. All of measured binaries are LMB. Hence, we assume an error caused by reddening is small enough because of relatively short distance between us and the binaries.}. For calibration of masses we used large tables of these indices on the website mentioned above. Here are the rules for deriving the mass and its uncertainty:
\begin{itemize}
	\item Write masses which correspond to the photometric indices (including an error\footnote{$J-H = 0.523 (59)$, write mass for $0.464$ and $0.582$}) using the tables. These two intervals (one for each index) are $I_1$ and $I_2$.
	\item 1. $I_3 = I_1 \cap I_2 = I_1$ or $I_3 = I_1 \cap I_2 = I_2$: then mass is average of $M_1$ and  $M_2$.
	\begin{equation}	
	M_1 = \left[\min(I_3)+\max(I_3)\right]/2
	\end{equation}
	\item 2. $I_1 \cap I_2 \neq I_1$ or $I_1 \cap I_2 \neq I_2$: then $\min(I_i)<\min(I_j)<\max(I_i)<\max(I_j)$. Then mass can be estimate as a weighted average where weights are lengths of the intervals:
	\begin{equation}
	\begin{split}
	M_1 &= \frac{\max(I_i)/\left[\max(I_i) - \min(I_i)\right]}{1/\left[\max(I_i) - \min(I_i)\right]+1/\left[\max(I_j) - \min(I_j)\right]} \\
	&+ \frac{\min(I_j)/\left[\max(I_j) - \min(I_j)\right]}{1/\left[\max(I_i) - \min(I_i)\right]+1/\left[\max(I_j) - \min(I_j)\right]}
	\end{split}
	\end{equation}
	\item 3. $I_1 \cap I_2 = \emptyset$: then $\min(I_i)<\max(I_i)<\min(I_j)<\max(I_j)$. Again, mass will be estimated as a weighted average:
	\begin{equation}
	\begin{split}
	M_1 &= \frac{\max(I_i)/\left[\max(I_i) - \min(I_i)\right]}{1/\left[\max(I_i) - \min(I_i)\right]+1/\left[\max(I_j) - \min(I_j)\right]} \\
	&+ \frac{\min(I_j)/\left[\max(I_j) - \min(I_j)\right]}{1/\left[\max(I_i) - \min(I_i)\right]+1/\left[\max(I_j) - \min(I_j)\right]}
	\end{split}
	\end{equation}
	\item Lower/Upper uncertainty was estimated similarly. For lower limit:

	\begin{equation}
		\begin{split}
	\sigma(M_1) &= \frac{\left|M_1 - \min(I_1)\right|/\left[\max(I_1) - \min(I_1)\right]}{1/\left[\max(I_1) - \min(I_2)\right] + 1/\left[\max(I_2) - \min(I_2)\right]}\\
	& +\frac{\left|M_1 - \min(I_2)\right|/\left[\max(I_2) - \min(I_2)\right]}{1/\left[\max(I_1) - \min(I_2)\right] + 1/\left[\max(I_2) - \min(I_2)\right]}
		\end{split}
	\end{equation}
	
	For upper limit there are $\max$ in absolute values instead of $\min$. Same rule was used for $\sigma(T)$ and $\sigma\left(M_{\mathrm{bol}}\right)$. In these cases we calculated average of lower and upper uncertainties.
	\item Mass of secondary component is computed as $M_2~=~M_1~\cdot~q$ and we used propagation of uncertainties (we assumed $\sigma(q) = 0.05$).
\end{itemize}

We deal with the effect, that seems to be caused by magnetic fields. Therefore, we must take into account the presence of stellar spots that deform the light curve. For the asymmetric light curves the stellar spots were considered as a hypothesis. However, the suitability of placing the spot(s) was tested using the so-called BIC criterion \citep{BIC} which connects the resulting chi-square with the number of parameters used and the degrees of freedom.

Firstly, we assume that the mass ratio is equal to 1. We then let the model parameters converge so that the theoretical curve coincides with the observed one. Because the mass ratio is probably not equal to 1, we used the M-L relation in the form
\begin{equation}
	\label{eq:mr}
	\log q = \frac{\log L_{\mathrm{bol}_2} - \log L_{\mathrm{bol}_1}}{4.841}
\end{equation}
where $L_{\mathrm{bol}_1}$ and $L_{\mathrm{bol}_2}$ are the bolometric luminosities. The equation was modified according to \citet{Graczyk_2003} and \citet{Eker_2015}. Then again we let the model parameters converge and the process is repeated until the mass ratio changes were negligible (less than 0.02) and the best $\chi^{2}$ value was obtained.

We describe the analysis of one system (NSVS~363024) step by step. Then some interesting systems were chosen and briefly mentioned in the following sections 4.1-4.7.

NSVS 363024 was observed between 31st Jan 2015 and 25th July 2016 (see Table \ref{tab:obs_info}). In this period we measured three primary and two secondary minima (Table \ref{tab:minima}). All images were processed in a standard way. We applied correction frames, chose a suitable comparison star and aperture. Then we obtained the light curves for each photometric filter. All the observed minima were used for refining the orbital period via an $O-C$ diagram analysis. Depending on the remoteness in time of observed minima, we have refined the orbital period of one to two orders of magnitude (compare values in Table \ref{tab:basic} and Table \ref{tab:parameters1}).

After that we estimate the mass and temperature of primary using the photometric indices (only $J-H$ and $H-K$ was known), as described above. We let the mass ratio equal to one and set reflection effect with 2 reflections. Fine and course grid raster for both components were set to 30. We started with hypothesis that NSVS 363024 is a detached binary system and chose this model as one of option of PHOEBE. We checked this hypothesis during the whole fitting process. All the available photometric filters were fitted simultaneously. At first, we fitted parameters that have a significant effect on the light curve. These are the relative luminosity of components $L_1$ and $L_2$, inclination $i$, surface gravity potentials $\Omega_1$ and $\Omega_2$ and temperature of secondary component $T_2$. Further, we let converge parameters of the second order (albedos $A_i$, coefficients of gravitational darkening $g_i$ and parameters of synchronicity $F_i$). They have considerable errors often, because our photometric observations were affected by observing conditions. Then we fitted some of first order parameters again and some of second order parameters but the values in the correlation matrix of these parameters should not exceed 0.7 and cannot exceed 0.8. This criterion selected groups of parameters which could be and were fitted together. Next, we can try to fit the third light $L_3$ and add a stellar spot. Spot's parameters could be estimated from asymmetry of the light curve. Because the effective temperature and radius of a spot are correlated we assumed that ratio of spot temperature and star temperature is higher than $0.65$ analogically to the sunspots. The ratio should by even higher for cooler stars \citep{Berdyugina_2005}. Finally, when the fit of the light curve is as good as it can be we change value of mass ratio as it is determined by Eq. \ref{eq:mr}. The whole procedure was repeated until the changes of the mass ratio are negligible. Then we can add more stellar spots and repeat the procedure. But we had to check the BIC criterion for optimal number of spots. Limb darkening coefficient was derived via tables published by \citet{Hamme_1993}. The whole fitting process was finished when $\chi^2$ stopped decreasing its value and residuals are symmetric around the zero value. A very similar process was applied to all systems.

We also try to fix the second order parameters ($F_i = 1$, $A_i = 0.6$ and $g_i = 0.32$) and carry out the analysis again. We did not change number of spots. The results are given in Table \ref{tab:RM_fix}. Changes of values of masses and radii are smaller than their uncertainties in many cases. It is the same for bolometric magnitudes and temperatures. There are two systems (NSVS 2517147 and NSVS 3630887) with significant change of radius (secondary components only). Their radius increase of about 0.1 $R_{\astrosun}$ and mass of the components increase of 0.04 $M_{\astrosun}$ and 0.07 $M_{\astrosun}$. The alternative solutions still give radii for these two binaries in disagreement with the models.

\section{Individual systems}
Table \ref{tab:basic} summarizes basic information about the studied systems and Table \ref{tab:obs_info} contains basic information about our observations. Some interesting systems were briefly described in the following subsections.

\begin{table*}
	\caption{Basic properties of systems from \citet{Hoffman_2008} and \citet{2MASS_2003} catalogs.}
	\label{tab:basic}
	\centering
    \begin{tabular}{p{2.8cm} c c c c c c}
		\hline\hline
		System & RA & DE & $J-H$ & $H-K$ & $V$ & $P$\\
		 & [h m s] & [$\circ$ $\prime$ $\prime\prime$] & [mag] & [mag] & [mag] & [days]\\
		\hline
		2MASS J01535799 +7146169 & 01 53 57.99 & $+$71 46 17.02 & 0.523 (59) & 0.099 (81) & - & -\\
		NSVS 363024 & 01 53 58.91 & $+$71 41 26.01 & 0.590 (36) & 0.114 (35) & 12.744 (41) & 0.43319\phantom{0}\\
		NSVS 401928 & 03 26 18.02 & $+$69 56 21.24 & 0.460 (37) & 0.163 (38) & 13.014 (52) & 0.57888\phantom{0}\\
		V0345 Cam & 04 25 55.35 & $+$69 15 45.51 & 0.545 (40) & 0.209 (40) & 13.361 (72) & 0.45157\phantom{0}\\
		NSVS 2285307 & 06 09 21.56 & $+$58 32 54.81 & 0.483 (81) & 0.256 (79) & 13.441 (81) & 0.64621\phantom{0}\\
		NSVS 2517147 & 09 19 31.31 & $+$57 17 30.24 & 0.477 (28) & 0.139 (26) & 13.048 (47) & 0.44673\phantom{0}\\
		GK CVn & 14 00 48.19 & $+$33 47 22.1\phantom{0} & 0.559 (29) & 0.104 (31) & $\sim$14.45\phantom{000\, (0)} & 0.401288\\
		V0514 Dra & 17 19 54.84 & $+$69 47 42.65 & 0.453 (25) & 0.075 (25) & 12.941 (69) & 0.31422\phantom{0}\\
		NSVS 3151384 & 20 19 49.09 & $+$65 34 14.28 & 0.620 (50) & 0.120 (40) & 11.305 (16) & 1.25079\phantom{0}\\
		2MASS J20333510 +4839427 & 20 33 35.10 & $+$48 39 42.73 & 0.248 (38) & 0.098 (38) & - & -\\
		NSVS 5789962 & 20 33 36.26 & $+$48 45 33.10 & 0.282 (37) & 0.044 (36) & 12.160 (38) & 1.45668\phantom{0}\\
		NSVS 3243815 & 20 38 57.73 & $+$58 04 56.74 & 0.510 (26) & 0.124 (29) & 13.177 (76) & 0.86394\phantom{0}\\
		NSVS 6127971 & 22 49 56.89 & $+$38 43 48.36 & 0.468 (78) & 0.153 (76) & 12.692 (41) & 0.54186\phantom{0}\\
		NSVS 3630887 & 23 55 39.53 & $+$39 12 10.62 & 0.611 (34) & 0.173 (32) & 12.439 (32) & 0.46030\phantom{0}\\
		\hline
	\end{tabular}
\end{table*}

\subsection{NSVS 363024}
\begin{figure}
	\centering
	\includegraphics[width=0.5\textwidth]{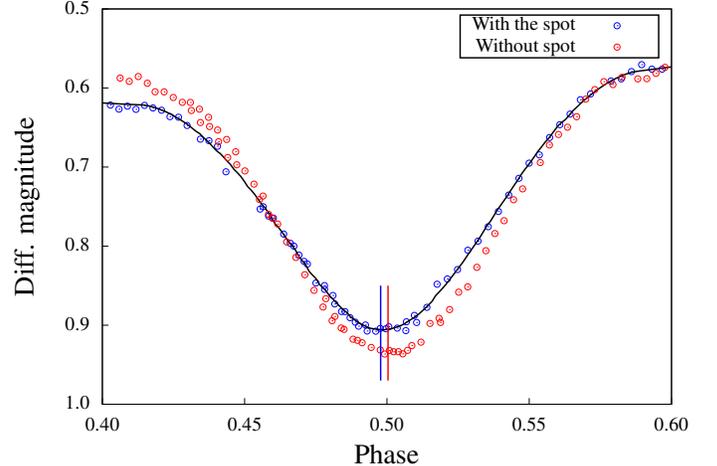}
	\caption{Secondary minimum of NSVS 363024. There is outlined a difference in mid-eclipse times in both cases. The difference is about 95 seconds. Times of minima were computed as minimum of a polynomial fitting of the direct and mirror image light curves.}
	\label{pic:spot_evol}
\end{figure}

NSVS 363024 (2MASS J01535891+7141260) is an eclipsing binary in Cassiopeia constellation. Its maximum brightness in filter V is 12.744 mag. There is not a single detailed publication about this binary, so the spectral type is unknown. It was estimated from the photometric indices. From the above described analysis it revealed that NSVS 363024 is a detached binary system. Light curve shows asymmetry, which was explained via a presence of a stellar spot. Half a year after the last measurement of the secondary minimum we measured that part of the curve twice again. Both observations were carried out in good weather conditions. The shape of the newly measured minima does not coincide with the shape of the minimum which was measured earlier (see Fig. \ref{pic:spot_evol} for comparison). Such a spot variability on the timescale of hundreds of days was observed on several other systems, see e.g. \citet{Skvrny}.

\subsection{V0345 Cam}
V0345 Cam (2MASS J04255534+6915455) is a detached system with its brightness of 13.361 mag in filter V. Two primary and three secondary minima were observed from 17 Apr 2015 to 19 Jan 2016. Both photometric indices gave us nearly the same masses and effective temperatures. Due to this fact only small uncertainties of masses and temperatures were calculated.

\subsection{NSVS 2285307}
NSVS 2285307 (2MASS J06092156+5832548) is a detached binary with the orbital period around 0.646 days. Its brightness in filter V is 13.441 mag. Data were carried out from 25 Feb 2015 to 09 Dec 2015 and we obtained three primary minima and two secondary minima. It allowed us to refined the orbital period about two orders of magnitude. The light curves of this system were asymmetric. This was explained by a spot on the surface of primary component. Its properties (and properties of others' spots) are in Table \ref{tab:spots}. Significant value of the third light (around 16 \%) was also detected. We did not find a star which is in the vicinity of the target which can be identified with the high level of the third light. It would be fruitful to carry out spectroscopy and test this hypothesis.

\subsection{V0514 Dra}
V0514 Dra (2MASS J17195483+6947426) is a W UMa type binary whose brightness in filter $\mathrm{V}_{\mathrm{max}}$ is 12.941 mag. We observed two primary and two secondary minima. On the light curve there is a slight asymmetry which was described by a hot spot on the surface of the primary component. This is the only hot spot as emerged from our analysis.

\subsection{NSVS 3151384}
NSVS 3151384 (2MASS J20194908+6534142) is a typical detached system. Its brightness in filter V is 11.305 mag. Likewise as in the previous case we monitored two primary and two secondary minima. We found very high value of a third light. It is due to a star which is very close to NSVS 3151384. Angular distance of these stars is less than 0.1 $\mathrm{arcmin}$.

\subsection{NSVS 3243815}
NSVS 324815 (2MASS J20385772+5804567) is another detached binary in our sample. Its light curves were slightly asymmetric (see Fig. \ref{pic:LCsall}) but we were not able to fit the curves with only one spot. We covered four primary minima and one secondary minimum. There was not any evolution of these spots in nearly two hundred days of the observational span.

\subsection{NSVS 3630887}
NSVS 3630887 (2MASS J23553953+3912106) is a detached binary. Maximum brightness of this system is 12.439 mag in filter V. We observed one primary and one secondary minimum and two incomplete secondary minima. The light curve of this system is strongly asymmetric. It was explained via only one spot on the surface of primary component. There was also detected a low value of the third light (only 2 \% of the luminosity). This small value could be caused by imprecise data.

\section{Results and discussion}
\begin{figure*}
	\centering
	\includegraphics[width=\textwidth]{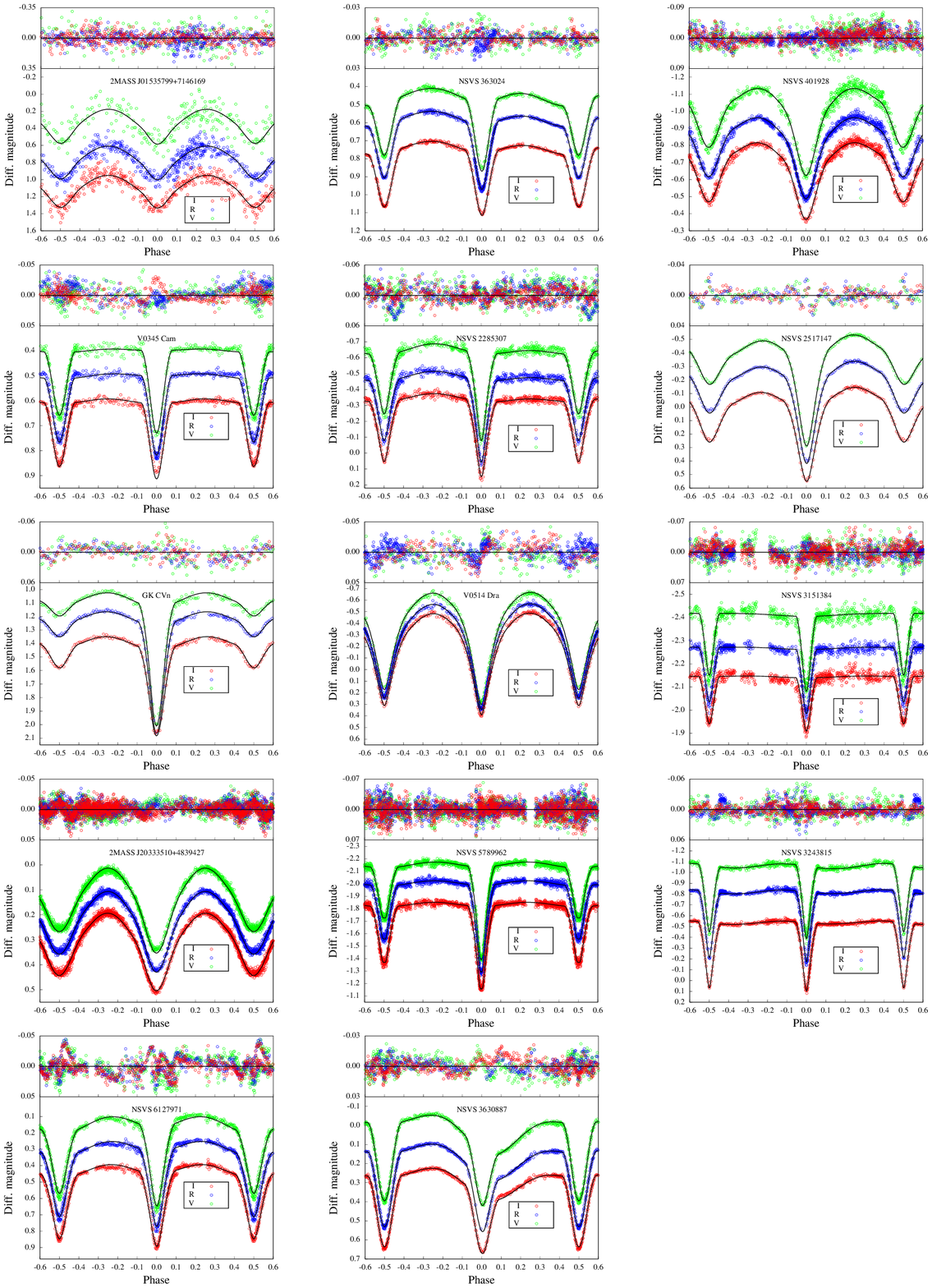}
	\caption{Light curves. Final fit according to PHOEBE.}
	\label{pic:LCsall}
\end{figure*}

The parameters of binaries we studied are summarized in the Tables \ref{tab:parameters1} and \ref{tab:parameters2}, their light curves are shown in Fig. \ref{pic:LCsall}. Properties of spots are listed in the Table \ref{tab:spots}. There are only values of luminosities in the filter R because the scatter of the data was the smallest and values of luminosity in other filters are nearly the same as in the filter R (the differences are few percent only). Mass ratio $q$ was calculated from Eq. (\ref{eq:mr}) and its values are from 0.84 to 1.00.

\begin{table*}
	\caption{Parameters of binaries (PHOEBE results).}
	\label{tab:parameters1}
	\centering
	\begin{tabular}{p{2.8cm} c c c c c c c c c}
		\hline\hline
		System & $H\!J\!D_0$ & $P$ & $i$ & $\Omega_1$ & $\Omega_2$ & L$_1$ & L$_2$ & L$_3$ & Spot\\
		 & [$-2400000$] & [days] & [deg] & & & [\%] & [\%] & [\%] & Y/N\\
		\hline
		2MASS J01535799 +7146169 & 57268.530\phantom{0} (52) & 0.27660\phantom{00} & 67.38 (71) & 3.71\phantom{0} (11) & 3.71\phantom{0} (11) & $\sim$50 & $\sim$50 & \phantom{~00}0 & N\\
		NSVS 363024 & 57268.4824 (42) & 0.4332050 & 77.06 (45) & 4.298 (57) & 4.643 (62) & $\sim$63 & $\sim$37 & \phantom{~00}0 & Y\\
		NSVS 401928 & 57117.4185 (31) & 0.5788356 & 71.09 (43) & 3.552 (37) & 3.636 (42) & $\sim$67 & $\sim$33 & \phantom{~00}0 & N\\
		V0345 Cam & 57297.4721 (37) & 0.451581\phantom{0} & 75.63 (51) & 4.781 (42) & 4.657 (36) & $\sim$55 & $\sim$45 & \phantom{~00}0 & N\\
		NSVS 2285307 & 57135.3155 (42) & 0.6462444 & 87.05 (92) & 4.761 (36) & 5.693 (39) & $\sim$51 & $\sim$33 & $\sim$16 & Y\\
		NSVS 2517147 & 57070.5801 (35) & 0.446696\phantom{0} & 78.56 (42) & 3.758 (29) & 3.841 (26) & $\sim$76 & $\sim$24 & \phantom{~00}0 & Y\\
		GK CVn & 57499.5681 (35) & 0.4012861 & 76.70 (26) & 4.856 (62) & 3.751 (73) & $\sim$75 & $\sim$25 & \phantom{~00}0 & N\\
		V0514 Dra & 57089.4184 (49) & 0.3142184 & 85.22 (76) & 3.536 (47) & 3.536 (47) & $\sim$50 & $\sim$50 & \phantom{~00}0 & Y\\
		NSVS 3151384 & 57206.4214 (41) & 1.250494\phantom{0} & 84.17 (93) & 7.047 (84) & 6.466 (75) & $\sim$30 & $\sim$32 & $\sim$38 & N\\
		2MASS J20333510 +4839427 & 57503.1674 (39) & 0.541725\phantom{0} & 62.91 (51) & 3.552 (41) & 3.552 (41) & $\sim$68 & $\sim$32 & \phantom{~00}0 & N\\
		NSVS 5789962 & 57294.4378 (51) & 1.456326\phantom{0} & 89.34 (86) & 4.974 (64) & 5.396 (72) & $\sim$68 & $\sim$32 & \phantom{~00}0 & N\\			
		NSVS 3243815 & 57238.8913 (37) & 0.864084\phantom{0} & 87.01 (62) & 6.402 (41) & 6.386 (72) & $\sim$52 & $\sim$48 & \phantom{~00}0 & Y\\
		NSVS 6127971 & 57296.4832 (21) & 0.541732\phantom{0} & 80.01 (55) & 4.203 (41) & 4.537 (43) & $\sim$60 & $\sim$40 & \phantom{~00}0 & N\\
		NSVS 3630887 & 57275.5392 (30) & 0.460195\phantom{0} & 78.62 (31) & 3.806 (37) & 5.070 (38) &	$\sim$67 & $\sim$31 & \phantom{0}$\sim$2 & Y\\
		\hline
	\end{tabular}
\end{table*}
\begin{table*}
	\caption{Spots properties.}
	\label{tab:spots}
	\centering
	\begin{tabular}{l c c c c c}
	\hline\hline
	System & Component & Colatitude & Longitude & Radius & Temperature\\
	& prim/sec & [deg] & [deg] & [deg] & $T_{\mathrm{spot}}/T_{\mathrm{photosphere}}$\\
	\hline
	NSVS 363024 & prim & 70 & 225 & 17 & 0.820\\
	NSVS 2285307 & prim & 38 & 277 & 20 & 0.724\\
	NSVS 2517147 & prim & 141 & 118 & 28 & 0.711\\
	V0514 Dra & prim & 139 & 209 & 18 & 1.113\\
	NSVS 3243815 & prim & 120 & 130 & 18 & 0.732\\
	& sec & 81 & 121 & 25 & 0.927\\
	NSVS 3630887 & sec & 92 & 134 & 33 & 0.650\\
	\hline
	\end{tabular}
\end{table*}
\begin{table*}
	\caption{Absolute parameters of binaries.}
	\label{tab:parameters2}
	\centering
	\begin{tabular}{p{2.8cm} c c c c c c c c}
		\hline\hline
		System & $M_1$ & $M_2$ & $R_1$ & $R_2$ & $M_{\mathrm{bol}_1}$ & $M_{\mathrm{bol}_2}$ & $T_1$ & $T_2$\\
		 & [$M_{\astrosun}$] & [$M_{\astrosun}$] & [$R_{\astrosun}$] & [$R_{\astrosun}$] & [mag] & [mag] & [K] & [K]\\
		\hline
		2MASS J01535799 +7146169 & 0.76 $\left(^{+12}_{-05}\right)$ & 0.76 $\left(^{+13}_{-06}\right)$ & 0.80 (08) & 0.79 (08) & 6.08\phantom{.} (78) & 6.09\phantom{.} (78) & 4750 (fixed) & 4750 (fixed)\\
		NSVS 363024 & 0.72 $\left(^{+07}_{-07}\right)$ & 0.66 $\left(^{+07}_{-07}\right)$ & 0.80 (04) & 0.69 (03) & 6.32\phantom{.} (72) & 6.81\phantom{.} (60) & 4500 (fixed) & 4309\phantom{..} (387)\\
		NSVS 401928 & 0.77 $\left(^{+06}_{-06}\right)$ & 0.67 $\left(^{+07}_{-07}\right)$ & 1.27 (09) & 1.13 (08) & 5.1\phantom{0} (1.1) & 5.72\phantom{0} (53) & 4750 (fixed) & 4334\phantom{..} (288)\\
		V0345 Cam & 0.44 $\left(^{+03}_{-01}\right)$ & 0.42 $\left(^{+03}_{-02}\right)$ & 0.62 (03) & 0.63 (03) & 7.95\phantom{.} (97) & 8.10\phantom{.} (83) & 3500 (fixed) & 3371\phantom{..} (36)\\
		NSVS 2285307 & 0.70 $\left(^{+14}_{-14}\right)$ & 0.64 $\left(^{+13}_{-13}\right)$ & 0.93 (06) & 0.77 (05) & 6.3\phantom{.} (1.0) & 6.8\phantom{.} (1.1) & 4200 (fixed) & 4140\phantom{..} (668)\\
		NSVS 2517147 & 0.80 $\left(^{+02}_{-02}\right)$ & 0.67 $\left(^{+04}_{-04}\right)$ & 1.02 (03) & 0.93 (03) & 5.46\phantom{.} (57) & 6.44\phantom{0} (48) & 4850 (fixed) & 4064\phantom{..} (111)\\
		GK CVn & 0.73 $\left(^{+15}_{-15}\right)$ & 0.67 $\left(^{+14}_{-14}\right)$ & 0.66 (05) & 0.89 (06) & 6.7\phantom{.} (1.1) & 7.1\phantom{.} (1.4) & 4550 (fixed) & 3547\phantom{..} (472)\\
		V0514 Dra & 0.76 $\left(^{+07}_{-07}\right)$ & 0.76 $\left(^{+08}_{-08}\right)$ & 0.89 (05) & 0.91 (05) & 5.85\phantom{.} (49) & 5.85\phantom{.} (49) & 4750 (fixed) & 4750 (fixed)\\
		NSVS 3151384 & 0.70 $\left(^{+08}_{-08}\right)$ & 0.70 $\left(^{+09}_{-09}\right)$ & 0.90 (07) & 1.00 (08) & 6.34\phantom{.} (95) & 6.2\phantom{.} (1.0) & 4200 (fixed) & 4090\phantom{..} (401)\\
		2MASS J20333510 +4839427 & 1.12 $\left(^{+10}_{-18}\right)$ & 0.96 $\left(^{+10}_{-17}\right)$ & 1.40 (09) & 1.31 (08) & 3.84\phantom{.} (66) & 4.64\phantom{.} (66) & 6000 (fixed) & 5167\phantom{..} (321)\\
		NSVS 5789962 & 1.04 $\left(^{+18}_{-07}\right)$ & 0.89 $\left(^{+16}_{-08}\right)$ & 1.66 (08) & 1.33 (07) & 3.7\phantom{.} (1.2) & 4.48\phantom{.} (61) & 5750 (fixed) & 5315\phantom{..} (322)\\
		NSVS 3243815 & 0.76 $\left(^{+04}_{-03}\right)$ & 0.75 $\left(^{+06}_{-05}\right)$ & 0.81 (03) & 0.80 (04) & 6.04\phantom{.} (36) & 6.13\phantom{.} (29) & 4750 (fixed) & 4690\phantom{..} (177)\\
		NSVS 6127971 & 0.80 $\left(^{+09}_{-12}\right)$ & 0.74 $\left(^{+10}_{-12}\right)$ & 1.00 (05) & 0.89 (05) & 5.50\phantom{.} (77) & 5.90\phantom{.} (73) & 4850 (fixed) & 4692\phantom{..} (475)\\
		NSVS 3630887 & 0.64 $\left(^{+06}_{-06}\right)$ & 0.54 $\left(^{+06}_{-06}\right)$ & 0.90 (04) & 0.56 (03) & 6.51\phantom{.} (99) & 7.44\phantom{.} (50) & 4050 (fixed) & 4157\phantom{..} (237)\\
		\hline
	\end{tabular}
\end{table*}

\begin{table*}
	\caption{Absolute parameters of binaries with fixed values of second order parameters.}
	\label{tab:RM_fix}
	\centering
	\begin{tabular}{p{2.8cm} c c c c c c c c}
		\hline\hline
		System & $M_1$ & $M_2$ & $R_1$ & $R_2$ & $M_{\mathrm{bol}_1}$ & $M_{\mathrm{bol}_2}$ & $T_1$ & $T_2$\\
		 & [$M_{\astrosun}$] & [$M_{\astrosun}$] & [$R_{\astrosun}$] & [$R_{\astrosun}$] & [mag] & [mag] & [K] & [K]\\
		\hline
		2MASS J01535799 +7146169 & 0.76 $\left(^{+12}_{-05}\right)$ & 0.76 $\left(^{+13}_{-06}\right)$ & 0.80 (08) & 0.80 (09) & 6.08\phantom{.} (78) & 6.08\phantom{.} (78) & 4750 (fixed) & 4750 (fixed)\\
		NSVS 363024 & 0.72 $\left(^{+07}_{-07}\right)$ & 0.67 $\left(^{+07}_{-07}\right)$ & 0.78 (04) & 0.73 (03) & 6.35\phantom{.} (70) & 6.73\phantom{.} (60) & 4500 (fixed) & 4277\phantom{..} (384)\\
		NSVS 401928 & 0.77 $\left(^{+06}_{-06}\right)$ & 0.69 $\left(^{+07}_{-07}\right)$ & 1.24 (09) & 1.18 (09) & 5.1\phantom{0} (1.1) & 5.61\phantom{0} (62) & 4750 (fixed) & 4350\phantom{..} (289)\\		
		V0345 Cam & 0.44 $\left(^{+03}_{-01}\right)$ & 0.42 $\left(^{+03}_{-02}\right)$ & 0.60 (03) & 0.59 (03) & 8.01\phantom{.} (92) & 8.26\phantom{.} (68) & 3500 (fixed) & 3345\phantom{..} (36)\\
		NSVS 2285307 & 0.70 $\left(^{+14}_{-14}\right)$ & 0.61 $\left(^{+13}_{-13}\right)$ & 0.99 (07) & 0.78 (05) & 6.2\phantom{.} (1.1) & 6.8\phantom{.} (1.1) & 4200 (fixed) & 4052\phantom{..} (654)\\	
		NSVS 2517147 & 0.80 $\left(^{+02}_{-02}\right)$ & 0.71 $\left(^{+04}_{-04}\right)$ & 0.98 (03) & 1.04 (03) & 5.56\phantom{.} (47) & 6.18\phantom{0} (24) & 4850 (fixed) & 4065\phantom{..} (111)\\	
		GK CVn & 0.73 $\left(^{+15}_{-15}\right)$ & 0.70 $\left(^{+15}_{-15}\right)$ & 0.63 (05) & 0.95 (07) & 6.8\phantom{.} (1.2) & 6.9\phantom{.} (1.3) & 4550 (fixed) & 3564\phantom{..} (474)\\
		V0514 Dra & 0.76 $\left(^{+07}_{-07}\right)$ & 0.76 $\left(^{+08}_{-08}\right)$ & 0.89 (05) & 0.89 (05) & 5.84\phantom{.} (50) & 5.84\phantom{.} (50) & 4750 (fixed) & 4750 (fixed)\\
		NSVS 3151384 & 0.70 $\left(^{+08}_{-08}\right)$ & 0.70 $\left(^{+09}_{-09}\right)$ & 0.91 (07) & 1.01 (08) & 6.33\phantom{.} (95) & 6.2\phantom{.} (1.0) & 4200 (fixed) & 4094\phantom{..} (401)\\
		2MASS J20333510 +4839427 & 1.12 $\left(^{+10}_{-18}\right)$ & 0.99 $\left(^{+10}_{-17}\right)$ & 1.40 (09) & 1.33 (08) & 3.85\phantom{.} (66) & 4.44\phantom{.} (66) & 6000 (fixed) & 5373\phantom{..} (333)\\
		NSVS 5789962 & 1.04 $\left(^{+18}_{-07}\right)$ & 0.89 $\left(^{+16}_{-08}\right)$ & 1.68 (08) & 1.35 (08) & 3.6\phantom{.} (1.2) & 4.44\phantom{.} (61) & 5750 (fixed) & 5342\phantom{..} (324)\\
		NSVS 3243815 & 0.76 $\left(^{+04}_{-03}\right)$ & 0.76 $\left(^{+06}_{-05}\right)$ & 0.77 (03) & 0.85 (04) & 6.17\phantom{.} (26) & 6.00\phantom{.} (39) & 4750 (fixed) & 4689\phantom{..} (177)\\
		NSVS 6127971 & 0.80 $\left(^{+09}_{-12}\right)$ & 0.74 $\left(^{+09}_{-12}\right)$ & 1.02 (05) & 0.87 (05) & 5.46\phantom{.} (80) & 5.90\phantom{.} (73) & 4850 (fixed) & 4754\phantom{..} (482)\\
		NSVS 3630887 & 0.64 $\left(^{+06}_{-06}\right)$ & 0.60 $\left(^{+07}_{-06}\right)$ & 0.85 (04) & 0.67 (03) & 6.64\phantom{.} (86) & 6.95\phantom{.} (57) & 4050 (fixed) & 4253\phantom{..} (245)\\
		\hline
	\end{tabular}
\end{table*}

\subsection{M-R relation}
\begin{figure*}
	\centering
	\includegraphics[width=0.95\textwidth]{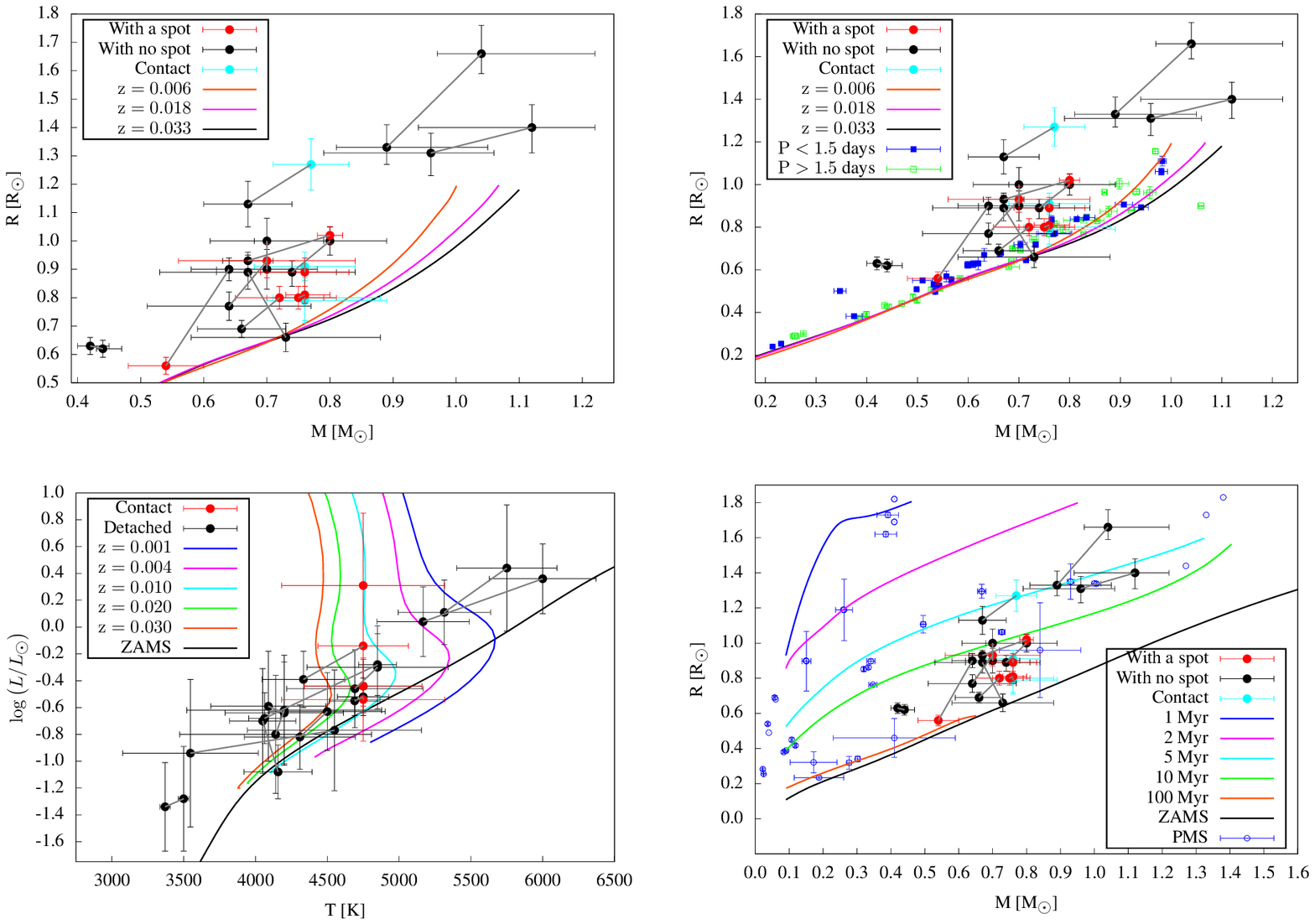}
	\caption{Top left: Masses and radii of measured stars. Components of a binary are connected by grey lines. The theoretical curves were calculated for five billion years old single stars with different values of metallicities.\\
	Top right: Comparison of our results with another systems for which photometry and spectroscopy were analysed simultaneously.\\
	Bottom left: H-R diagram. Models were computed via Evolve ZAMS code \citep{Paxton_EvolveZAMS} for $M = 0.6\,M_{\astrosun}$.\\
	Bottom right: M-R relation for PMS stars. Models were adopted from \citet{PMS}.}
	\label{pic:RMplusHR}
\end{figure*}

In Fig. \ref{pic:RMplusHR} (top left) there are masses and radii of measured stars. Grey lines connect the components of one binary. The curves in the figure are theoretically calculated M-R relation for single stars evolution for five billion years after ZAMS for three different values of metallicity \citep{Spada_2013}. The values of radii are independent on metallicity for masses lower than approximately $0.7\,M_{\astrosun}$. Hence, we did not take any assumptions of metallicities. Although the theoretical curves are calculated for the given age, it is a good approximation for LMB due to their slow evolution. But the difference caused by evolution can be significant for solar-type stars (four of the studied components).

Only one of the measured star is not located above the theoretical curves of the M-R relation. The discrepancy is not more evident for components with the spots on the surface. For five systems there is a component with and a component without the spot. For three of these systems spot-covered component is further from the curve but in two cases it is the opposite case. Uncertainties of masses of primary components are from 3 \% to 20 \% and secondary components from 7 \% to 21 \%. In only few cases the accuracy is good enough to study M-R discrepancy. Hence, it seems that sole photometry cannot be a suitable method for studying the M-R relation discrepancy. In Fig. \ref{pic:RMplusHR} (top right) our results are compared with masses and radii of LMB for which photometry and spectroscopy were analysed simultaneously (a list of these components is given in Table \ref{tab:RM}).

\subsection{Another explanation of discrepancy}
We can ask if the result will be different when we had a better photometric data. Masses of stars are independent of our data due to the photometric indices calibration. Uncertainties of radii could be smaller, if we had more accurate data but probably the discrepancy would remain.

What if measured stars are old enough and their radii are larger due to evolution of stars? To disprove this possibility we calculated an evolution of stars with different masses and metallicities via Evolve ZAMS code \citep{Paxton_EvolveZAMS}. The result was that the discrepancy could not be explained due to the evolution: stars have to be older than the Universe. We also verified that radius is independent on metallicity. Different metallicity have an impact on H-R diagram (see Fig. \ref{pic:RMplusHR} bottom left). For explanation of stars' positions on H-R diagram we would need young stars with high metallicity. The uncertainties of temperatures and brightness are big enough to say final statement.

What if these stars have not reached the main sequence yet? They could have bigger radii due to the fact that they are not yet in the hydrostatic equilibrium. In Fig. \ref{pic:RMplusHR} (bottom right) there are M-R relations for pre-main sequence stars \citep{PMS}. This could be a possible explanation which is a very unlikely explanation because the measured stars are not in any nebula, they are ordinary stars in sky field. There are only a few known pre-main sequence eclipsing binaries and it should be an extremely lucky coincidence. Hence to conclude, without spectroscopy we cannot disprove or confirm any such hypothesis.

Another possibility could be unsuitable mass calibration. We tested whether the change of an initial mass of the primary component will have a significant impact on the entire result. This discrepancy could be explained only if the stars have much higher masses. This can be explained by interstellar extinction. It causes that our estimates of masses are more likely lower estimates. On the other hand, these stars are very likely to be in our neighbourhood and that the difference should not be so substantial and IR filters are not so affected by interstellar extinction. To conclude, the problem with the mass calibration is an open question.

\section{Conclusion}
Masses and radii are shown at Fig. \ref{pic:RMplusHR} (top left). Only one of the studied stars is not above the theoretical curve of the M-R relation. We did not observe difference between the radius of the stars that are spotted and those unspotted ones. It is not in accordance with the interpretation of \citet{Chabrier_2007} but due to large uncertainties we cannot say anything about this difference. The uncertainties are up to 21 \% in determining the masses. The improvement of accuracy could be achieved by using more robust calibration. We compared our results with a compilation of 66 components of binaries where both spectroscopy and photometry were analysed simultaneously (see Fig. \ref{pic:RMplusHR} top right). Another explanation of discrepancy in our data may be very low stellar ages (between 5 Myr to 100 Myr) as is shown in Fig.~\ref{pic:RMplusHR} (bottom right).

Due to the large uncertainties in determining the masses it seems that only photometric observations could not be suitable method for studying the different current models of stellar evolution and observed parameters of LMB. It would be very useful to obtain a good set of spectra of the systems studied and verify the accuracy of the calibration of masses.

\begin{acknowledgements}
We are grateful to K. Hornoch for the smooth operation of the telescope. The authors would like to thank the anonymous referees for their valuable comments and suggestions to improve the quality of the paper. This investigation was supported by the Czech Science Foundation grant No. GA15-02112S. This research has made use of the SIMBAD and VIZIER databases, operated at CDS, Strasbourg, France and of NASA’s Astrophysics Data System Bibliographic Services.
\end{acknowledgements}

\bibliographystyle{apj}
\bibliography{BIBL}

\clearpage
\clearpage

\onecolumngrid
\begin{longtable}{l c c c c}

\caption{Masses, radii and orbital periods of double-lined low-mass binary components which uncertainties of masses and radii are less than 5 \%.}
\label{tab:RM}\\
\hline\hline
System & Mass & Radius & Period & References\\
& [$M_{\astrosun}$] & [$R_{\astrosun}$] & [d] &\\
\hline
\endfirsthead
\multicolumn{5}{c}{\tablename\ \thetable{}: Continued}\\
\hline\hline
System & Mass & Radius & Period & References\\
& [$M_{\astrosun}$] & [$R_{\astrosun}$] & [d] &\\
\hline
\endhead
\hline
\endfoot

\hline
\endlastfoot
BX Tri A & 0.51 (2) & 0.55 (1) & 0.192 & 1\\
NSVS 01031772 A & 0.5428 (27) & 0.5260 (28) & 0.368 & 2\\
NSVS 01031772 B & 0.4982 (25) & 0.5088 (30) & 0.368 & 2\\
NGC2204-S892 A & 0.733 (5) & 0.719 (14) & 0.452 & 3\\
NGC2204-S892 B & 0.662 (5) & 0.680 (17) & 0.452 & 3\\
GU Boo A & 0.6101 (64) & 0.627 (16) & 0.489 & 4, 5\\
GU Boo B & 0.5995 (64) & 0.624 (16) & 0.489 & 4, 5\\
NSVS 02502726 A & 0.714 (19) & 0.645 (6) & 0.560 & 6\\
NSVS 02502726 B & 0.347 (12) & 0.501 (5) & 0.560 & 6\\
RT And B & 0.907 (17) & 0.906 (11) & 0.629 & 4, 7\\
MG1-1819499 A & 0.557 (1) & 0.569 (25) & 0.630 & 8\\
MG1-1819499 B & 0.535 (1) & 0.500 (17) & 0.630 & 8\\
CG Cyg A & 0.941 (14) & 0.893 (12) & 0.631 & 4, 7\\
CG Cyg B & 0.814 (13) & 0.838 (11) & 0.631 & 4, 7\\
GSC 08814-01026 A & 0.833 (17) & 0.845 (12) & 0.702 & 1\\
GSC 08814-01026 B & 0.703 (13) & 0.718 (17) & 0.702 & 1\\
GJ 3236 A & 0.375 (16) & 0.3829 (57) & 0.771 & 9\\
YY Gem A & 0.5992 (47) & 0.6194 (57) & 0.814 & 4, 10\\
YY Gem B & 0.5992 (47) & 0.6194 (57) & 0.814 & 4, 10\\
MG1-116309 A & 0.567 (2) & 0.552 (17) & 0.827 & 8\\
MG1-116309 B & 0.532 (2) & 0.532 (12) & 0.827 & 8\\
UV Psc A & 0.9829 (77) & 1.110 (23) & 0.861 & 4, 11\\
UV Psc B & 0.7644 (45) & 0.835 (18) & 0.861 & 4, 11\\
GSC 04894-02310 A & 0.771 (33) & 0.772 (12) & 0.897 & 1\\
GSC 04894-02310 B & 0.768 (33) & 0.769 (13) & 0.897 & 1\\
AE For A & 0.6314 (35) & 0.67 (3) & 0.918 & 12\\
AE For B & 0.6197 (34) & 0.63 (3) & 0.918 & 12\\
CM Dra A & 0.23102 (89) & 0.2534 (19) & 1.268 & 4, 13\\
CM Dra B & 0.21409 (83) & 0.2398 (18) & 1.268 & 4, 13\\
IM Vir A & 0.981 (12) & 1.061 (16) & 1.309 & 14\\
IM Vir B & 0.6644 (48) & 0.681 (13) & 1.309 & 14\\
BD-15 2429 A & 0.7029 (45) & 0.694 (9) & 1.528 & 1\\
BD-15 2429 B & 0.6872 (49) & 0.699 (13) & 1.528 & 1\\
MG1-506664 A & 0.584 (2) & 0.560 (5) & 1.548 & 8\\
MG1-506664 B & 0.544 (2) & 0.513 (9) & 1.548 & 8\\
MG1-78457 A & 0.527 (2) & 0.505 (15) & 1.586 & 8\\
MG1-78457 B & 0.491 (1) & 0.471 (16) & 1.586 & 8\\
TYC 4749-560-1 A & 0.8338 (36) & 0.848 (5) & 1.622 & 1\\
TYC 4749-560-1 B & 0.8280 (40) & 0.833 (5) & 1.622 & 1\\
MG1-646680 A & 0.499 (2) & 0.457 (10) & 1.638 & 8\\
MG1-646680 B & 0.443 (2) & 0.427 (8) & 1.638 & 8\\
MG1-2056316 A & 0.469 (2) & 0.441 (4) & 1.723 & 8\\
MG1-2056316 B & 0.382 (1) & 0.374 (4) & 1.723 & 8\\
RX J0239.1-1028 A & 0.7300 (90) & 0.7410 (40) & 2.072 & 15, 16\\
RX J0239.1-1028 B & 0.6930 (60) & 0.7030 (20) & 2.072 & 15, 16\\
FL Lyr B & 0.958 (12) & 0.962 (28) & 2.178 & 4, 17\\
ZZ UMa B & 0.9691 (48) & 1.1562 (96) & 2.299 & 4\\
V1061 Cyg Ab & 0.9315 (68) & 0.974 (20) & 2.347 & 4, 18\\
V1236 Tau A & 0.787 (12) & 0.788 (15) & 2.588 & 19\\
V1236 Tau B & 0.770 (9) & 0.817 (10) & 2.588 & 19\\
CU Cnc A & 0.4349 (12) & 0.4323 (55) & 2.771 & 4, 20\\
CU Cnc B & 0.39922 (89) & 0.3916 (94) & 2.771 & 4, 20\\
G 179-55 A & 0.2576 (85) & 0.2895 (68) & 3.550 & 9, 21\\
G 179-55 B & 0.2585 (80) & 0.2895 (68) & 3.550 & 9, 21\\
T-Cyg1-12664 A & 0.680 (21) & 0.613 (7) & 4.129 & 1\\
V636 Cen B & 0.8545 (30) & 0.8300 (43) & 4.284 & 4\\
V818 Tau A & 1.0591 (62) & 0.900 (16) & 5.609 & 10\\
V818 Tau B & 0.7605 (62) & 0.768 (10) & 5.609 & 10\\
HS Aur A & 0.898 (19) & 1.004 (24) & 9.815 & 4, 17\\
HS Aur B & 0.877 (17) & 0.874 (24) & 9.815 & 4, 17\\
RW Lac B & 0.8688 (40) & 0.9638 (40) & 10.369 & 4, 22\\
V568 Lyr B & 0.8273 (42) & 0.7679 (64) & 14.470 & 4\\
KIC 6131659 A & 0.922 (7) & 0.8800 (28) & 17.528 & 23\\
KIC 6131659 B & 0.685 (5) & 0.6395 (61) & 17.528 & 23\\
FBS 1109+767 A & 0.3951 (22) & 0.3814 (31) & 41.032 & 24\\
FBS 1109+767 B & 0.2749 (11) & 0.3001 (45) & 41.032 & 24\\
\end{longtable}
%%%
\begin{footnotesize}
\textbf{References:} 1 \citet{Cakirli_BX}, 2 \citet{Lopez_NSVS}, 3 \citet{Rozyczka_NGC}, 4 \citet{Torres_2010}, 5 \citet{Lopez_GU}, 6 \citet{Cakirli_NSVS}, 7 \citet{Popper_CG_RT}, 8 \citet{Kraus_MG1}, 9 \citet{Irwin_GJ}, 10 \citet{Torres_YY_V818}, 11 \citet{Popper_UV}, 12 \citet{Rozyczka_AE}, 13 \citet{Morales_CM}, 14 \citet{Morales_IM}, 15 \citet{Lopez_RX}, 16 \citet{Feiden_2012}, 17 \citet{Popper_FL_HS}, 18 \citet{Torres_V1061}, 19 \citet{Bayless_V1236}, 20 \citet{Ribas_CU}, 21 \citet{Hartman_1RXS}, 22 \citet{Lacy_RW}, 23 \citet{Bass_KIC}, 24 \citet{Irwin_LSPM}
\end{footnotesize}

%%%%%%%%%%%%%%%%%%%%%%%%%%%%% ONLINE %%%%%%%%%%%%%%%%%%%%%%%%%%%%
%%%%%%%%%%%%%%%%%%%%%%%%%%%%%%%%%%%%%%%%%%%%%%%%%%%%%%%%%%%%%%%%%
\begin{appendix}

\section{Observation log}

\begin{table*}
	\caption{Information about observations.}
	\label{tab:obs_info}
	\centering
    \begin{tabular}{l c c c c c}
		\hline\hline
		System & From & To & $N$ & Aperture & Comparison star\\
		 & [d/m/y] & [d/m/y] & [I/R/V] & [arc sec] &\\
		\hline
		2MASS J01535799+7146169 & 31/01/15 & 25/07/16 & 375/483/208 & 1.4 & 2MASS J01535745+7146059\\
		NSVS 363024 & 31/01/15 & 25/07/16 & 295$^a$/435$^a$/307$^a$ & 3.7 & 2MASS J01545478+7139110\\
		NSVS 401928 & 04/04/15 & 17/08/16 & 620/800/606 & 3.7 & 2MASS J03261195+6956176\\
		V0345 Cam & 17/04/15 & 19/01/16 & 317/358/311 & 3.7 & 2MASS J04263015+6917324\\
		NSVS 2285307 & 25/02/15 & 09/12/15 & 517/559/551 & 3.7 & 2MASS J06083785+5830425\\
		NSVS 2517147 & 30/01/15 & 16/02/15 & 131/135/133 & 4.9 & 2MASS J09195939+5711028\\
		GK CVn & 20/04/16 & 15/09/16 & 135/125/136 & 2.8 & 2MASS J14011012+3346159\\
		V0514 Dra & 07/03/15 & 03/06/15 & 219/485/226 & 3.7 & 2MASS J17210185+6953376\\
		NSVS 3151384 & 10/05/15 & 19/01/16 & 660/648/686 & 7.7 & 2MASS J20195374+6536570\\
		2MASS J20333510+4839427 & 28/09/15 & 13/11/16 & 1514/1359/1393 & 2.8 & 2MASS J20332212+4843435\\
		NSVS 5789962 & 28/09/15 & 08/11/16 & 1264/1226/1303 & 3.7 & 2MASS J20334670+4846449\\
		NSVS 3243815 & 26/02/15 & 26/01/16 & 530/650/542 & 2.8 & 2MASS J20382276+5802417\\
		NSVS 6127971 & 24/09/15 & 17/08/16 & 445/440/455 & 3.7 & 2MASS J22501858+3843502\\
		NSVS 3630887 & 02/09/15 & 05/10/15 & 413/353/409 & 4.9 & 2MASS J23554892+3907373\\
		\hline
		\multicolumn{6}{l}{$^a$ Does not contain data with evolved spot.}
	\end{tabular}
\end{table*}

\section{Summary of observations}

\begin{longtable*}{l l l l}
%\captionsetup{width=\textwidth}
\caption{Summary of observations. Exp. time in format I/R/V (zero means we do not have any data from this filter).}
\label{tab:summary}\\
\hline\hline
System & Date & Exp. time & Weather condition\\
& & [s] &\\
\hline
\endfirsthead
\multicolumn{4}{c}{\tablename\ \thetable{}: Continued}\\
\hline\hline
System & Date & Exp. time & Weather condition\\
& [d/m/y] & [s] &\\
\hline
\endhead
\hline
\endfoot
\hline
\multicolumn{4}{l}{$^a$ The data were used for deriving the mid-eclipse time.}
\endlastfoot
2MASS J01535799+7146169 & 31/01/15 & 0/30/0 & cirrus\\
& 27/08/15 & 30/30/30 & clouds all the time, Moon\\
& 02/09/15 & 30/30/30 & clouds, Moon, high humidity, fog\\
& 04/09/15 & 45/30/45 & clouds, Moon\\
& 19/01/16 & 30/30/30 & clear, Moon\\
& 24/02/16 & 30/30/30 & clear\\
& 25/07/16 & 30/30/30 & some clouds, Moon\\
\hline
NSVS 363024 & 31/01/15 & 0/30/0 & cirrus\\
& 27/08/15 & 30/30/30 & clouds all the time, Moon\\
& 02/09/15 & 30/30/30 & clouds, Moon, high humidity, fog\\
& 04/09/15 & 45/30/45 & clouds, Moon\\
& 19/01/16 & 30/30/30 & clear, Moon\\
& 24/02/16 & 30/30/30 & clear\\
& 25/07/16 & 30/30/30 & some clouds, Moon\\
\hline
NSVS 401928 & 04/04/15 & 0/60/0 & clear\\
& 18/11/15 & 30/30/30 & some clouds, windy\\
& 09/12/15 & 30/30/30 & clouds, high humidity\\
& 19/01/16 & 30/30/30 & clouds, Moon\\
& 26/01/16 & 30/30/30 & clouds, Moon, high humidity\\
& 10/02/16 & 30/30/30 & clouds\\
& 24/02/16 & 30/30/30 & clouds, Moon\\
& 25/05/16 & 30/30/30 & clouds\\
& 18/08/16 & 30/30/30 & clear, Moon\\
\hline
V0345 Cam & 17/04/15 & 0/45/0 & clear\\
& 04/09/15 & 45/45/45 & clouds, Moon\\
& 01/10/15 & 45/45/45 & clear, Moon\\
& 24/10/15 & 60/60/60 & cirrus, Moon\\
& 03/11/15 & 30/20/40 & mostly clear, Moon\\
& 19/01/16 & 30/20/40 & clouds, Moon\\
\hline
NSVS 2285307 & 26/11/14 & 0/60/90 & some clouds\\
& 25/02/15 & 60/60/60 & clouds, Moon\\
& 02/03/15 & 60/60/60 & clouds, Moon\\
& 16/03/15 & 60/60/60 & clouds\\
& 13/04/15 & 60/60/60 & somewhat cloudy\\
& 22/04/15 & 60/60/60 & some clouds\\
& 27/10/15 & 60/60/60 & clear, Moon\\
& 03/11/15 & 45/30/45 & clear\\
& 18/11/15 & 45/30/45 & clouds, windy\\
& 24/11/15 & 45/30/45 & thin cirrus\\
& 09/12/15 & 45/30/45 & clear, very high humidity\\
\hline
NSVS 2517147 & 30/01/15 & 0/60/0 & clouds$^a$\\
& 16/02/15 & 60/60/60 & clear\\
\hline
GK CVn & 20/04/16 & 80/80/80 & clear, Moon\\
& 28/08/16 & 80/80/80 & thin cirrus\\
& 15/09/16 & 0/60/0 & clear, Moon\\
\hline
V0514 Dra & 07/03/15 & 0/30/0 & clear\\
& 13/04/15 & 30/30/30 & somewhat cloudy\\
& 29/04/15 & 0/30/0 & cirrus, Moon\\
& 04/05/15 & 30/30/30 & clouds, Moon\\
& 03/06/15 & 30/30/30 & clear\\
\hline
NSVS 3151384 & 10/05/15 & 0/60/0 & clear\\
& 03/06/15 & 45/45/45 & clear\\
& 02/07/15 & 45/45/45 & clear, Moon\\
& 29/07/15 & 45/45/45 & somewhat cloudy, Moon\\
& 30/07/15 & 45/45/45 & some clouds, Moon\\
& 31/07/15 & 45/45/45 & clouds, Moon\\
& 03/08/15 & 45/45/45 & clear, Moon\\
& 05/08/15 & 45/30/45 & clouds, Moon\\
& 27/10/15 & 45/30/45 & clear, Moon\\
& 03/11/15 & 30/30/45 & clear\\
& 19/01/16 & 30/30/45 & clouds\\
\hline
2MASS J20333510+4839427 & 28/09/15 & 30/30/30 & clouds, Moon\\
& 24/11/15 & 30/30/30 & clear, Moon\\
& 24/04/16 & 30/30/30 & some clouds, Moon\\
& 25/05/16 & 30/30/30 & clear, Moon\\
& 06/06/16 & 30/30/30 & clear\\
& 21/06/16 & 30/30/30 & clear, Moon\\
& 29/06/16 & 30/20/30 & clear\\
& 25/07/16 & 30/20/30 & some clouds\\
& 31/08/16 & 30/20/30 & clear\\
& 15/09/16 & 30/20/30 & clear, Moon\\
& 13/11/16 & 30/20/30 & clear, Moon\\
\hline
NSVS 5789962 & 28/09/15 & 30/30/30 & clouds, Moon\\
& 24/11/15 & 30/30/30 & clear, Moon\\
& 24/04/16 & 30/30/30 & some clouds, Moon\\
& 25/05/16 & 30/30/30 & clear, Moon\\
& 06/06/16 & 30/30/30 & clear\\
& 21/06/16 & 30/30/30 & clear, Moon\\
& 29/06/16 & 30/20/30 & clear\\
& 25/07/16 & 30/20/30 & some clouds\\
& 31/08/16 & 30/20/30 & clear\\
& 15/09/16 & 30/20/30 & clear, Moon\\
\hline
NSVS 3243815 & 26/02/15 & 0/45/0 & clear, Moon\\
& 10/04/15 & 0/45/0 & clear\\
& 03/07/15 & 45/45/45 & clear\\
& 04/07/15 & 45/45/45 & cirrus\\
& 07/07/15 & 45/45/45 & clear\\
& 03/08/15 & 45/45/45 & clear, Moon\\
& 25/08/15 & 45/45/45 & clouds, later clear, Moon\\
& 27/08/15 & 45/45/45 & clouds, Moon\\
& 20/09/15 & 45/45/45 & clouds\\
& 24/11/15 & 45/45/45 & thin cirrus, Moon\\
& 19/01/16 & 45/45/45 & clouds, Moon\\
& 26/01/16 & 45/45/45 & some clouds, Moon\\
\hline
NSVS 6127971 & 24/09/15 & 45/45/45 & clouds, Moon\\
& 30/09/15 & 45/45/60 & clear, Moon\\
& 18/11/15 & 30/30/30 & clouds, windy\\
& 19/01/16 & 30/30/30 & clear, Moon\\
& 06/06/16 & 30/30/30 & clear\\
& 29/61/16 & 30/30/30 & clear\\
& 25/07/16 & 30/30/30 & some clouds\\
& 17/08/16 & 30/30/30 & clear, Moon\\
\hline
NSVS 3630887 & 02/09/15 & 30/30/30 & clouds, high humidity, Moon, fog\\
& 09/09/15 & 30/30/30 & high humidity\\
& 20/09/15 & 30/30/30 & clouds\\
& 05/10/15 & 45/45/45 & clouds\\
\end{longtable*}

\section{Tables of minima}

\begin{longtable*}{l c c c}
%\captionsetup{width=\textwidth}
\caption{Times of mid-eclipses. Ephemerides were adopted from Table \ref{tab:parameters1}.}
\label{tab:minima}\\
\hline\hline
System & Time of eclipse & p/s & Epoch\\
& [HJD-2400000] & & \\
\hline
\endfirsthead
\multicolumn{4}{c}{\tablename\ \thetable{}: Continued}\\
\hline\hline
System & Time of eclipse & p/s & Epoch\\
& [HJD-2400000] & &\\
\hline
\endhead
\hline
\endfoot
\hline
\endlastfoot
		2MASS J01535799+7146169 & 57054.450 (34) & p & -774.0\\
		& 57262.3092 (31) & s & -22.5\\
		& 57262.5842 (22) & s & -21.5\\
		& 57268.5285 (19) & p & 0.0\\
		& 57270.4681 (28) & p & 7.0\\
		& 57443.3381 (43) & p & 632.0\\
		& 57595.612 (14) & s & 1182.5\\
		\hline
		NSVS 363024 & 57054.47924 (50) & p & -494.0\\
		& 57262.6333 (75) & p & -13.5\\
		& 57268.48295 (19) & p & 0.0\\
		& 57270.43109 (26) & s & 4.5\\
		& 57407.3253 (96) & s & 320.5\\
		& 57443.2815 (16) & s & 403.5\\
		& 57595.5523 (16) & p & 755.0\\
		\hline
		NSVS 401928 & 57117.41825 (23) & p & 0.0\\
		& 57345.1854 (29) & s & 393.5\\
		& 57407.4091 (25) & p & 501.0\\
		& 57414.36165 (99) & p & 513.0\\
		& 57414.6500 (17) & s & 513.5\\
		& 57429.4133 (28) & p & 539.0\\
		& 57443.30108 (77) & p & 563.0\\
		& 57443.5921 (13) & s & 563.5\\
		& 57618.40194 (26) & s & 865.5\\
		\hline
		V0345 Cam & 57130.3872 (12) & p & -370.0\\
		& 57270.60216 (90) & s & -59.5\\
		& 57297.47183 (17) & p & 0.0\\
		& 57320.2770 (55) & s & 50.5\\
		& 57320.5016 (45) & p & 51.0\\				
		& 57330.66337 (23) & s & 73.5\\
		& 57407.43340 (62) & s & 243.5\\
		\hline
		NSVS 2285307 & 56988.2942 (25) & s & -277.5\\
		& 57079.4148 (11) & s & -86.5\\
		& 57084.2634 (11) & p & -79.0\\
		& 57135.31620 (42) & p & 0.0\\
		& 57165.3655 (18) & s & 46.5\\
		& 57330.48189 (26) & p & 302.0\\
		& 57351.48411 (19) & s & 334.5\\
		& 57366.6703 (22) & p & 358.0\\
		\hline
		NSVS 2517147 & 57053.3827 (73) & s & -38.5\\
		& 57070.35638 (33) & s & -0.5\\
		& 57070.58140 (20) & p & 0.0\\
		\hline
		GK CVn & 57499.36503 (23) & s & -0.5\\
		& 57499.56797 (19) & p & 0.0\\
		& 57647.2404 (12) & p & 368.0\\
		\hline
		V0514 Dra & 57089.41835 (34) & p & 0.0\\
		& 57126.49623 (76) & p & 118.0\\
		& 57126.65323 (87) & s & 118.5\\
		& 57142.36430 (17) & s & 168.5\\
		& 57147.39192 (54) & s & 184.5\\
		& 57177.39940 (39) & p & 280\\
		\hline
		NSVS 3151384 & 57206.42061 (23) & p & 0.0\\
		& 57233.3027 (39) & s & 21.5\\
		& 57234.55858 (40) & s & 22.5\\
		& 57323.34199 (53) & s & 93.5\\
		& 57330.22016 (30) & p & 99.0\\
		\hline
		2MASS J20333510+4839427 & 57294.33255 (89) & s & -385.5\\
		& 57351.2105 (44) & s & -280.5\\
		& 57503.4376 (10) & s & 0.5\\
		& 57546.5019 (27) & p & 80.0\\
		& 57561.40109 (55) & s & 107.5\\
		& 57569.5273 (24) & s & 122.5\\
		& 57595.5280 (31) & s & 170.5\\
		& 57632.36678 (66) & s & 238.5\\
		& 57632.6349 (22) & p & 239.0\\
		& 57647.5358 (12) & s & 266.5\\
		& 57706.31349 (69) & p & 375.0\\
		\hline
		NSVS 5789962 & 57294.43858 (30) & p & 0.0\\
		& 57351.2336 (41) & p & 39.0\\
		& 57503.42106 (29) & s & 143.5\\
		& 57546.38177 (16) & p & 173.0\\
		& 57632.30643 (10) & p & 232.0\\
		& 57647.59699 (26) & s & 242.5\\
		\hline
		NSVS 3243815 & 57207.3531 (70) & s & -36.5\\
		& 57238.46067 (28) & s & -0.5\\		
		& 57260.49427 (34) & p & 25.0\\
		& 57286.41699 (30) & p & 55.0\\
		& 57351.22274 (95) & p & 130.0\\
		& 57407.3884 (56) & p & 195.0\\
		& 57414.3018 (12) & p & 203.0\\
		\hline
		NSVS 6127971 & 57290.52442 (93) & p & -11.0\\
		& 57296.48408 (66) & p & 0.0\\
		& 57407.26738 (39) & s & 204.5\\
		& 57546.49165 (22) & s & 461.5\\
		& 57569.51647 (30) & p & 504.0\\
		& 57618.54307 (41) & s & 594.5\\
		\hline
		NSVS 3630887 & 57268.40597 (56) & s & -15.5\\
		& 57275.30876 (20) & s & -0.5\\
		& 57275.54046 (25) & p & 0.0\\
		& 57286.35337 (26) & s & 23.5\\
		& 57301.5403 (13) & s & 56.5\\
\end{longtable*}
\end{appendix}

\end{document}